\documentclass[conference]{IEEEtran}
\usepackage{cite}
\usepackage{graphicx}
\graphicspath{ {images/} }
\usepackage{listings}
\usepackage{makecell}
\usepackage[T1]{fontenc}
\usepackage{wrapfig}
\usepackage[format=plain, belowskip=-15pt,aboveskip=-2pt,labelfont=it,textfont=it]{caption}
\newcommand{\rom}[1]{\MakeUppercase{\romannumeral #1}}

\begin{document}

%Here goes the title

\title{Driving asynchronous distributed tasks with events}

%Authors List

\author
{\IEEEauthorblockN{Nick Brown}
\IEEEauthorblockA{EPCC, \\The University of Edinburgh\\ Email: n.brown@epcc.ed.ac.uk}
\and
\IEEEauthorblockN{Oliver Thomson Brown}
\IEEEauthorblockA{EPCC, \\The University of Edinburgh}
\and 
\IEEEauthorblockN{J. Mark Bull}
\IEEEauthorblockA{EPCC, \\The University of Edinburgh}
}
\maketitle

%Main body starts

\begin{abstract}
Open-source matters, not just to the current cohort of HPC users but also to potential new HPC communities, such as machine learning, themselves often rooted in open-source. Many of these potential new workloads are, by their very nature, far more asynchronous and unpredictable than traditional HPC codes and open-source solutions must be found to enable new communities of developers to easily take advantage of large scale parallel machines. Task-based models have the potential to help here, but many of these either entirely abstract the user from the distributed nature of their code, placing emphasis on the runtime to make important decisions concerning scheduling and locality, or require the programmer to explicitly combine their task-based code with a distributed memory technology such as MPI, which adds considerable complexity. In this paper we describe a new approach where the programmer still splits their code up into distinct tasks, but is explicitly aware of the distributed nature of the machine and drives interactions between tasks via events. This provides the best of both worlds; the programmer is able to direct important aspects of parallelism whilst still being abstracted from the low level mechanism of how this parallelism is achieved. We demonstrate our approach via two use-cases, the Graph500 BFS benchmark and in-situ data analytics of MONC, an atmospheric model. For both applications we demonstrate considerably improved performance at large core counts and the result of this work is an approach and open-source library which is readily applicable to a wide range of codes.
\end{abstract}

\section{Introduction}
More and more, HPC is demonstrating to new communities the benefits it can provide. From machine learning and data analytics to interactive supercomputing, these represent important new application areas that heavily rely on open-source software and can benefit greatly from the power of supercomputing. Whilst traditional HPC users were often more willing to accept closed source tools such as compilers and libraries, these new generations of application areas have often grown up in the open-source community and as such expect a level of openness that HPC has not required before. But these new use-cases bring with them their own challenges and issues, for instance many of these new are not naturally regular and the traditional approach of writing bulk synchronous parallel codes, that arguably MPI has excelled at over the past 25 years, is no longer applicable. 
With a focus on using open-source to build a complete system from the hardware upwards, it is not just the applications that are driving future challenges in parallel programming. There are a generation of new open-source HPC hardware technologies, such as the RISC-V \cite{waterman2014risc} architecture, which will likely mean that future open-source systems will expose significant amounts of parallelism that the programmer must exploit. In order to be able to take advantage of these future machines it will be very important for more traditional computationally based applications to scale to many hundreds of thousands, if not millions, of cores, and to do this one needs to break apart the bulk synchronous nature of these codes.

It is our view that programming models which encourage and support the programmer to write highly asynchronous codes will be of great use in the future. We actually see this sort of approach already in other fields, for instance web developers leverage AJAX to make asynchronous calls from a website to a server and \emph{at some point} their callback will be executed with the resulting data. In the mean time their code can be executing other functionality and this ability to work out of lock-step avoids excessive idle time and maintains user interactivity. However, not least because web-browsers tend to only run a single thread, HPC is far more complex than web-browsers and a key question is how best to achieve this whilst abstracting the programmer from all the low level concerns of parallelism which they must currently face with technologies like MPI. It is already fairly common for developers of asynchronous applications to write their own, bespoke, communication layer, but this is not ideal and they are forced to do this due to the inadequacies of the programming technologies. Instead, following the philosophy of the open-source community, this should be abstracted away in a reusable library such that developers are free to concentrate on their application, with the underlying technology efficiently and seamlessly supporting the asynchronous nature of their parallel code.

Task-based models are seen as a possible solution, where the programmer splits up the computation into distinct tasks that are eligible for execution once a number of dependencies have been met. A big benefit of this is that the underlying mechanism of parallelism is abstracted, so that the programmer can concentrate on their application without worrying about the tricky low level details. Task-based models also encourage programmers to break their code into loosely coupled, distinct, components which promotes asynchronicity. However, a major limitation is that these models have traditionally been grounded in a shared memory assumption which significantly restricts the problem sizes that they can address. Whilst there is effort under way to apply task-based models to distributed memory machines, many still maintain this shared memory abstraction, which places great emphasis on the runtime to make the best decisions in matters impacting parallelism or require the programmer to explicitly mix distributed memory technologies such as MPI.

% Especially suited for irregular applications (our task model) - last part of next paragraph needs some thought

In this paper we present a new, open-source, approach to task-based models, where the programmer still works with the concept of tasks but is explicitly aware of the distributed nature of their code. It is our belief that, by making explicit the distribution, programmers can gain the best of both worlds - an ability to architect their parallelism based upon a realistic view of how this will execute so as to optimise performance, complete with a task-based model abstracting them from the low level details of how this interaction takes place. It is our hypothesis that this encourages and supports the programmer to write highly asynchronous codes, where the developer can set up interactions but does not need to worry about exactly when these will take place in their code.

This paper is organised as follows; Section \rom{2} describes our approach in detail and provides a simple code example to further flesh out the concepts. In Section \rom{3} we discuss related work currently being undertaken by the community, and compare and contrast this with our approach. Section \rom{4} introduces more advanced aspects of our approach to ensure generality to a wide range of applications. To give the reader a flavour of the generality of our approach we demonstrate two use-cases in sections \rom{5} and \rom{6}. An EDAT implementation of the BFS kernel from the Graph500 benchmark is studied in Section \rom{5} and the in-situ data analytics of an atmospheric model, MONC, is explored as a short case-study in Section \rom{6}. Lastly, in Section \rom{7} we draw conclusions and discuss further work.

% Good for highly asynchronous workloads - why needed? As want things that interact irregularly without making code very complex
% Effective approach is to follow the "when stuff completes then handle it"
% Dont think it is strong enough yet on the async and interoperability - why use threads too? <-- my view of the asynchronicty here!

\section{Event Driven Asynchronous Tasks (EDAT)}
\label{sec:edatoverview}
% can be thought of as combining AM with task based models, but simpler!
Our approach is known as Event Driven Asynchronous Tasks (EDAT) and a library has been developed to implement and test these concepts. The idea is to provide the programmer with a realistic view of the distributed nature of their code but to still abstract them from the low level details of how data is communicated by using a task-based model. The programmer is still working with the view of multiple nodes, each containing a distinct memory space and a number of CPU cores that can execute code. Hence in this paper we will use the terminology of process, a process being a distinct memory space running over one or more cores (such as a NUMA region) and these contain a number of \emph{workers}. There are two key concepts in EDAT:

\begin{itemize}
  \item \textbf{Tasks} are represented by the programmer as a function and are submitted to EDAT for running locally once a number of (event) dependencies have been met. The submission of a task is non-blocking, and once the dependencies have been met the task is eligible to run and moved into a \emph{ready queue} for execution by a \emph{worker} when one becomes available. It is common (and the default behaviour) to map one \emph{worker} per CPU core, but this is configurable by the user. Tasks can be nested, where a task submits other tasks.
  \item \textbf{Events} are the major way in which tasks and processes interact. An event is fired from one process to another (which can be the same process) and are labelled with an event identifier (EID). Events may contain some optional payload data which the task that consumes it is then able to process. The firing of events is non-blocking, and data itself is copied into an internal buffer such that once an event has been fired, the programmer can modify the data later on in their code without any worry of inconsistency.
\end{itemize}

Whilst events might seem similar to messages, crucially the way in which events are sent and delivered to tasks is entirely abstracted from the programmer. The programmer need not care about the low level details of how events are physically sent, handling events when they arrive, how these are mapped to the dependencies of tasks, or how tasks are queued up. Programmers work with the concept of tasks, but are explicitly aware of the distributed nature of their application and drive interactions through events. 

% No strict ordering, i.e. events can arrive or tasks be scheduled first, it doesnt matter

% Interoperability of threading (shared memory) and distributed memory - really easy and implicit here, not so when explicitly mixing MPI+openmp etc....

\subsection{Submission of tasks}
Tasks are submitted to EDAT for execution by the programmer and this can be done at any point in the code. Listing \ref{lst:submittask} illustrates the API call, where \emph{task\_function\_ptr} is a function pointer to the task itself. The \emph{num\_event\_dependencies} argument is the number of event dependencies that must be met before this task is eligible for execution. After this argument, for each event, the programmer provides the source rank (process ID) of where that event originates from, and a string event identifier which is depicted as \emph{$<$int event\_source, char * event\_identifier$>$} in Listing \ref{lst:submittask}. The programmer can view this as a task consuming events that match the task's dependencies.

\begin{lstlisting}[frame=lines,caption={Submitting a task to EDAT},label={lst:submittask}]
void edatSubmitTask(task_function_ptr, int num_event_dependencies, <int event_source, char * event_identifier>...)
\end{lstlisting}

Listing \ref{lst:taskfunction} illustrates the function signature of a task, where both the number of events (\emph{number\_of\_events}) that a task has been executed with, and an array of structures containing each event, \emph{events}, are passed to the task function as arguments. The \emph{EDAT\_Event} structure contains the optional payload data and also metadata about the event, including the source rank, string event identifier, data type of payload data and number of elements in the payload data. The library guarantees that the order in which events are delivered to a task matches the dependency order that the programmer expressed when submitting the task. Based on this the programmer can make assumptions about the order of events in the task's array argument and need not search through event metadata, which could slow down tasks with large numbers of dependencies.

\begin{lstlisting}[frame=lines,caption={Task function},label={lst:taskfunction}]
void task(EDAT_Event * events, int number_of_events) {
   ....
}
\end{lstlisting}

EDAT also provides some inbuilt constants that the programmer can use for an event's source argument. For instance \emph{EDAT\_SELF} specifies that the event should originate from the same rank and \emph{EDAT\_ANY} is a wild-card where the task dependency will be met based upon an event with a matching string identifier from any source process. 

% Terminology - tasks consume events, I don't think I really phrase it like this and it would be good to do so
\subsection{Firing of events}
% Return int from these! - what this for?
% Ordering behaviour (of arrival)
Events are fired from a source process to a target process, and this firing of events can occur at any point in an application. Events serve two purposes; firstly they contribute to the activation of tasks by matching dependencies of those tasks; and secondly they provide the ability to communicate data between tasks if required. We promote the notion of \emph{fire and forget} on behalf of the programmer, so the EDAT library makes a copy of any payload data during this call, and the user's code will continue to execute regardless of whether the event has been delivered, or of the state of the corresponding task. The programmer can immediately reuse the data passed to the event, and once fired there is no way for the programmer to detect the status of an event or cancel it. Instead there is simply the guarantee that fired events will, at some point, be delivered to the target rank.

\begin{lstlisting}[frame=lines,caption={Firing an event},label={lst:fireevent}]
void edatFireEvent(void* data, int data_type, int number_elements, int target_rank, const char * event_identifier)
\end{lstlisting}

Listing \ref{lst:fireevent} illustrates the API call for firing an event where \emph{data} is optional payload data to send, \emph{data\_type} is the type of this data (EDAT provides in-built types including integers, booleans, floats and doubles), and \emph{number\_elements} the number of elements contained in the data. The event will be fired to the \emph{target\_rank} (which can be \emph{EDAT\_SELF}) and the string \emph{event\_identifier} is the identifier used to match up this event with dependencies of outstanding tasks.

The ordering of events and task dependencies is deterministic and EDAT maintains a strict ordering of events fired from a target to source process. This guarantees that events will arrive at a target process in the same order in which they fired from a source process. The order in which tasks consume events is determined by the order in which they are submitted to EDAT; a task submitted before another task on a process has a higher precedence in the consumption of events.

\subsection{Bringing these ideas together: A simple example}
In designing EDAT it was our intention to keep the concepts and API as simple as possible, and the calls described so far are sufficient to develop an irregular parallel application. Listing \ref{lst:simpleexample} illustrates a simple parallel EDAT code running with (at least) two processes. As can be seen the programmer typically writes their code SPMD style and EDAT provides a \emph{edatGetRank} call to obtain the current rank which can be used to specialise which processes execute what code. In this example, at line 7, Process 0 submits the task, \emph{task1}, which has no event dependencies (the second argument given to \emph{edatSubmitTask} is zero) and so this task will be immediately eligible for execution by an idle worker on that process. 

Process 1 submits two tasks, \emph{task2} at line 9 which depends on a single event from Process 0 with event identifier \emph{event1}. \emph{task3} is submitted at line 10 with two event dependencies, an event from Process 0 with identifier \emph{event2} and an event from Process 1 with the identifier \emph{event3}.

As \emph{task1} (lines 16 to 20) executes on Process 0, two events are fired.  The first event, fired to Process 1 with the identifier \emph{event1} at line 17 contains no payload data (\emph{EDAT\_NONE} as the datatype). This event corresponds to the dependency of \emph{task2} on Process 1 and as such \emph{task2} will be eligible for execution by an idle worker as soon as this event arrives. The second event fired by \emph{task1} to Process 1, line 19, contains a single integer as payload data and this event corresponds to the first dependency of \emph{task3}. The other dependency of \emph{task3} is fired at line 24 by \emph{task2} running on Process 1 to itself (note the \emph{EDAT\_SELF} target.) Once both these events have arrived on Process 1 then the dependencies of \emph{task3} have been met and this task will be eligible for execution. The task itself extracts out the payload data from each event, adds these integers together, and prints the result.

\begin{lstlisting}[frame=lines,caption={Simple EDAT example code},label={lst:simpleexample},numbers=left, framexleftmargin=1.5em, xleftmargin=2em]
#include "edat.h"
#include <stdio.h>

int main() {
  edatInit();
  if (edatGetRank() == 0) {
    edatSubmitTask(task1, 0);
  } else if (edatGetRank() == 1) {
    edatSubmitTask(task2, 1, 0, "event1");
    edatSubmitTask(task3, 2, 0, "event2", 1, "event3");
  }
  edatFinalise();
  return 0;
}

void task1(EDAT_Event * events, int num_events) {
  edatFireEvent(NULL, 0, EDAT_NONE, 1, "event1");
  int d=33;
  edatFireEvent(&d, 1, EDAT_INT, 1, "event2");
}

void task2(EDAT_Event * events, int num_events) {
  int d=100;
  edatFireEvent(&d, 1, EDAT_INT, EDAT_SELF, "event3");
}

void task3(EDAT_Event * events, int num_events) {
  printf("%d\n", *((int*) events[0].data) + *((int*) events[1].data));
}
\end{lstlisting}

\begin{figure}[htb]
\centering\includegraphics[width=0.25\textwidth]{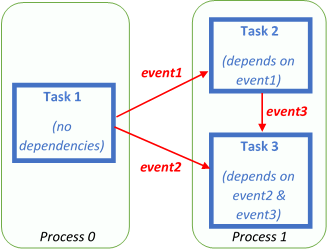}\caption{Diagrammatic task interaction of simple example}

\label{fig:diagramsimpleexample}
\end{figure}

Figure \ref{fig:diagramsimpleexample} illustrates these same tasks and interactions of the code from Listing \ref{lst:simpleexample}. The programmer determines what interacts with what and based upon this they don't need to worry about how events are physically delivered or when, just that they will and tasks depending upon these events will execute when they arrive. It can be seen from this example that our approach encourages the programmer to write asynchronous, loosely coupled, code as there is no explicit synchronisation or assumptions when events will arrive. Instead the programmer is submitting tasks based upon a number of events arriving in the knowledge that, at some point, these tasks will execute but they need not care about exactly when or impose limitations on the progress of other processes.

If one was to try and achieve the same thing in a technology like MPI directly then they would need to address many more low level concerns, such as explicitly testing for messages, carrying around numerous request handles, explicitly mapping functions to threads and repeatedly checking on progress. 

Hence in our approach the programmer is much more of an architect of parallelism; they are designing what aspects of their code interact, but not how this actually happens. As many of the performance issues with codes are caused by a poor high level design \cite{mattson2004patterns} we believe that this approach of supporting the programmer writing distributed, asynchronous codes in a manner that they can concentrate on the high level design is valuable. Given enough tasks, the overhead of sending many small messages, which can be unavoidable with irregular applications, can also be mitigated with an asynchronous task-based approach such as ours because message latency can be effectively hidden.

% Promote asynchronicity as can schedule stuff and it will run at some point when the events arrive. Can chain things via events too

% Omitted error checking on type, number of elements and number of events for brevity

% Simple as unlike things like Charm++ still think in terms of simple interactions and a general model that they are likely already familiar with

\subsection{Collective dependencies and events}
\label{sec:collectivedependencies}
The interactions described until now have been point to point, sent from a distinct source process to a distinct target process. EDAT also provides the \emph{EDAT\_ALL} constant that can be used for source and/or target ranks. When used as the target rank in firing events it will send that events to all processes (i.e. a broadcast), when used as a dependency in submitting tasks this requires events from all processes in order to meet that dependency. Listing \ref{lst:reductiontasks} sketches out a naive reduction in EDAT, where a task is submitted to run when an event with identifier \emph{event} is received from all processes (including itself.) Once this dependency is met the task will be mapped to an idle worker for execution and in the example here the task sums up integers from each process (in the payload data of each event.) Whilst in reality the programmer would most likely adopt a more complex collective algorithm, such as a tree-based approach, this would work equally well with EDAT and the simple example here is for illustration.

\begin{lstlisting}[frame=lines,caption={Reduction tasks},label={lst:reductiontasks}]
edatSubmitTask(task, 1, EDAT_ALL, "event");

void task(EDAT_Event * events, int num_events) {
  int total=0;
  for (int i=0;i<num_events;i++) {
  	total+=*((int*) events[i].data);
  }
}
\end{lstlisting}

By combining collective events with collective task dependencies, one can build a barrier as illustrated in listing \ref{lst:barrier}. Because the submission of tasks and firing of events is non-blocking, crucially this is a non-blocking barrier, and the task will not execute on a worker of any process until all processes have fired their corresponding events and these have been received. Therefore the execution of this task represents that each process has reached some logical point in the code by all processes. One use for this is when interacting with external, blocking collective calls (such as parallel NetCDF) where when individual processes call a function they block until all other processes have issued a corresponding call. This can result in excessive wait time, so instead an asynchronous barrier can be used at this call point and the barrier task is then executed at similar times on each process (assuming there are idle workers) which calls the blocking collective function and effectively reduces the time a process is blocking for other processes to issue the corresponding call.

\begin{lstlisting}[frame=lines,caption={Barriers in EDAT},label={lst:barrier}]
edatSubmitTask(task, 1, EDAT_ALL, "event");
.....
edatFireEvent(NULL, EDAT_NONE, 0, EDAT_ALL, "event");
\end{lstlisting}

\subsection{Termination}
The \emph{edatFinalise} call of listing \ref{lst:simpleexample} will put the main thread to sleep and wait until the code has finished executing before terminating the program. Four conditions must be met on all processes before EDAT terminates:
\begin{enumerate}
\item There are no outstanding submitted tasks waiting for their dependencies to be satisfied
\item All tasks eligible for execution have been full executed and all workers are now idle
\item All events have been delivered to their target process
\item All events have been consumed by tasks (i.e. there are no outstanding events waiting to satisfy the dependencies of a task not yet submitted).
\end{enumerate}

Once these four conditions are met on every process then EDAT will terminate. Of course the programmer is entirely abstracted from the mechanism of this, but should be aware that submitted tasks need to run and events need to be consumed before their code can terminate. 

\subsection{Implementation}
We wrote the EDAT library in C++ and the library itself is fairly small, at around 4000 lines of code. The scheduler follows a First In First Out (FIFO) policy for executing tasks, which are queued up when they are submitted. The underlying transport layer is designed to be pluggable, currently, due to its ubiquity, there is an MPI based implementation and other mechanisms can be easily added. Progress polling for aspects such as new events and termination can either be carried out by a dedicated thread or an idle worker. In the later case, polling is mapped to an idle worker but this worker prioritises task execution and polling will be \emph{swapped out} in preference to running a task if there are no other idle workers available.

%When the library is first initialised each worker is mapped to a thread and the affinity of that thread is set to the appropriate CPU core. Tasks are executed by workers, and if the programmer chooses to pause a task (see section \ref{sec:pausetasks}) then that thread is put to sleep via a condition variable. A new thread is then started, the correct affinity for the worker set and this worker is now able to execute new tasks on the same core. When a paused task is reactivated, the sleeping thread is added to a \emph{ready} queue and then reactivated when a worker becomes idle, the affinity of this reactivated thread set to that of the appropriate worker and this replaces the idle worker's previous thread. To avoid excessive thread creation and destruction when a thread is finished with, it is often stored for future re-use. This one of the reasons why we use the terminology of \emph{worker} instead of \emph{thread}; because, whilst each worker is running via a thread, there might be additional threads inside the runtime which are paused or stored for re-use and we did not want to overload the term.

\section{Related work}
As a community, many of our parallel programming tools, such as MPI, are best suited to applications where the parallelism is predictable and regular. In cases where interactions and application behaviour is itself unpredictable and irregular, then these existing tools start to come unstuck. The asynchronicity of codes that is now required by these developments means that existing paradigms, whilst they have served us well in a regular, synchronous world, are more and more becoming a hindrance to the programmer and are not suited to the communication patterns of a growing number of applications. The reader can view our approach as a combination of the event based co-ordination pattern \cite{mattson2004patterns}, with concepts from the task-based models paradigm to support the programming of asynchronous, large scale, parallel codes.

\subsection{Task-based models}
One possible solution to programming irregular applications are task-based models and typically these models commonly present themselves to the programmer via pragmas or a library API. Technologies such as OpenMP tasks \cite{openmp} and OmpSs \cite{fernandez2014task} require the programmer to annotate their code with pragmas which are then processed by the compiler. In these approaches the programmer expresses their dependencies based upon input and output variables and the compiler and runtime is able to generate a task dependency graph based upon this information. A task with specific input dependencies can not run until other tasks with these same dependencies as outputs have executed. In this manner the programmer is splitting their code up into these distinct, loosely coupled, work units and directing the order in which they can run based upon their variables (including function arguments.) 

Alternatively the model presents itself entirely as an API, whether it be a function library or other abstraction such as C++ templates. Legion \cite{slaughterlegion}, High Performance ParalleX (HPX) \cite{HPX} and the Open Community Runtime (OCR) \cite{mattson2016open} are examples of this approach. One benefit here is that there is no need for a specialist compiler, but many of these library approaches require the programmer to learn significant new concepts and techniques, before undertaking large scale code re-factoring. Fundamentally this is because with the pragma approach there is a compiler that can extract out sufficient information, whereas with libraries this information must be supplied directly by the programmer at runtime, and technologies such as Legion result in fairly verbose code. Whilst EDAT is a library based approach, we have been careful to limit the number of new concepts and API calls that the programmer need learn to effectively use the technology. 

Irrespective of how the task model it is presented, interoperability between shared memory and distributed memory technologies is a challenge when one wishes to scale task-based codes. There is a significant increase in complexity when one addresses distributed memory architecture, including challenges of optimal data placement, efficient communication and task locality. Broadly, there are two common approaches; either to place the burden of these challenges onto the runtime and keep it simple for the programmer with the programming interface largely unchanged (for instance OmpSs has adopted a directory/cache\cite{bueno2015run}), or place the burden on the programmer and require them to explicitly mix the task-based and distributed memory (such as MPI) technologies. 

Whilst the nature of pragma based approaches means that they invariably heavily rely on the concept of variables in their abstraction (which makes moving away from a shared memory abstraction challenging), library based models also commonly hide any distribution. For instance OCR \cite{mattson2016open}, an asynchronous task-based runtime, has the notation of event-driven tasks. However their concept of event driven tasks and our approach is very different. OCR abstracts data migration across distributed memory spaces via data blocks which contain an event that maps to task dependencies. Crucially here, though, the programmer has limited control and visibility over the locality of these data blocks, not least because the placement of OCR tasks, which can only generate one single event, is entirely abstracted by the runtime. This lack of transparency is compounded by the fact that OCR is not intended to be used directly by the application developer but instead as a lower level building block supporting higher level task-based models. This lack of visibility for the programmer, where it is not necessarily obvious how their data is distributed and the cost of communication related to this, limits opportunities for optimisation. Indeed EDAT demonstrated superior performance and scaling in contrast to OCR up to 32768 cores with Intel's stencil 2D benchmark in a earlier work \cite{edatposter}. 

\subsection{Active messaging}

Another attractive approach to abstracting asynchronicity from the programmer is that of active messaging \cite{von1992active}, where a sender provides not only the data but also a message handler to be executed by the receiver once the message arrives. In HPC the main success of this approach has been at the lower level transport layer, such as GASNet\cite{bonachea2002gasnet}, instead of technologies that are designed to be accessible to the application developer.

A number of active message end user libraries have been developed, for instance AM++ \cite{willcock2010am++} and AMMPI \cite{ammpi}, but these have deficiencies which limits their utility for irregular applications and uptake by the community has been poor. For instance AM++ is bulk synchronous in operation and communications take place only in epochs. Neither technologies utilise background threads for monitoring the sending or receiving of messages and the user has to handle this manually (either by calling the poll function directly, creating a thread to do this or in AM++ when the epoch ends.) These active messaging technologies are not multi-threaded and as such there is no model to execute message handlers concurrently.

To some extent the reader can view our approach as a combination of task-based models with some of the ideas from active messaging. Unlike the active messaging technologies, our approach is more abstract where EDAT events contribute towards the activation of code and can be composed together before a task will run. Charm++ \cite{kale1996charm++} is potentially closest to our approach where the programmer writes their code in C++ and defines parallel objects, called chares. Conceptually these chares are objects and contain a number of entry methods that can be invoked remotely by calling methods on proxy objects on other cores. There are a number of key differences between our approach and Charm++:
\begin{itemize}
\item In Charm++ the programmer must write their code Object Oriented (OO) style in C++ with extra boilerplate \emph{proxies}. This often requires completely rethinking and redesigning their application in OO C++. Charm++ is more of an ecosystem, rather than a task-based model, and as such there is no way of gradually adopting Charm++. Instead codes needs to be ported in a \emph{big-bang approach}, this is not the case for EDAT which is language agnostic.
\item In following the OO model, Charm++ limits the programmer to sending data from one process to another process when instantiating a remote method (effectively a task). It is possible to get round this with a Charm++ \emph{reduction}, but that sits outside the basic abstraction and adds extra complexity. Instead, in EDAT, events from any number of different processes can contribute to the execution of a task (the task dependencies), it's that simple.
\item The Charm++ model provides a default \emph{one-at-a-time} execution model, where only one method of an object can be executing at any one time. It is possible to override this but then the programmer is on their own. In EDAT we take the opposite view, and whilst it is possible to structure the submission of tasks for this \emph{one-at-a-time} approach this is not imposed. Additionally EDAT provides finer grained synchronisation controls, which will be described in section \ref{sec:sharingdata} integrated with EDAT to provide richer control.
\item There is no way in Charm++ to drip feed data to remote methods (tasks) and as such all communication must take place before a method is invoked. Instead in EDAT, as we will see in section \ref{sec:pausetasks}, it is possible to start tasks based upon a minimal number of dependencies and then test or wait for the arrival of further events. This provides more flexibility to the programmer in controlling task granularity.
\item Whilst the Charm++ source code is available, the licence is much more limiting than those more commonly associated with open-source software. For instance commercial use of Charm++ requires the purchase of a licence, any user modified version must be clearly marked and renamed to notify users that it is not the original software, users can only re-use 10\% of the source code without obtaining a further licence and code developed using Charm++ must cite specific papers. In short, whilst the source code of Charm++ is freely available the licence restrictions means that it feels a long way from the freedoms enjoyed by open-source software such as our own EDAT library.
\end{itemize}

By bringing in the ideas and concepts from the task-based paradigm and driving tasks with events we believe that this achieves the broad aims of active messaging but in a clear and unambiguous manner. 

\section{Advanced EDAT concepts}
Whilst the techniques described in Section \ref{sec:edatoverview} are sufficient for many applications, a number of more advanced concepts are useful for specific applications. 
\subsection{Persistent tasks and events}
\label{sec:persistent}
The tasks and events discussed up until this point are transitory. This means that when a task's dependencies are met it is then removed from the \emph{outstanding} queue and hence the task only executes once. Likewise; when transitory events are consumed, then they are gone. However, it is also possible for both tasks and events to be persistent, where tasks can run any number of times and events consumed any number of times.

\begin{lstlisting}[frame=lines,caption={Submitting persistent tasks to EDAT},label={lst:persistent_tasks}]
void edatSubmitPersistentTask(task_function_ptr, int num_event_dependencies, <int event_source, char * event_identifier>)
\end{lstlisting}

Listing \ref{lst:persistent_tasks} illustrates the API call for persistent tasks which contains identical arguments to the standard \emph{edatSubmitTask} for transitory tasks, just the call has a different name. This means the programmer can submit a persistent task which will be executed whenever the event dependencies are met. In the case of multiple task dependencies there can be multiple versions of a persistent task in progress, where some dependencies have been met for a persistent task a number of times and other dependencies are still outstanding for each copy of that persistent task. It is also possible to associate a name with persistent tasks so that programmers can then manage them later on in code, for example to remove them. Unlike transient tasks, persistent tasks are not required to have executed before the code can terminate. 

\begin{lstlisting}[frame=lines,caption={Firing a persistent event},label={lst:persistent_events}]
void edatFirePersistentEvent(void* data, int data_type, int number_elements, int target_rank, const char * event_identifier)
\end{lstlisting}

Listing \ref{lst:persistent_events} denotes the API call for persistent events which, as soon as they have been consumed as a task dependency will be fired again. However this re-firing of the event is done locally; even if a persistent event is sent from a remote process then the fact it is persistent it handled by the target.

\subsection{Pausing tasks and checking for events}
\label{sec:pausetasks}
In an ideal world one would write their tasks in a way that only requires interaction via their input dependencies. However we found that more flexible support was useful. The first of these, a task \emph{wait}, is illustrated in listing \ref{lst:taskwait} where at some point in the task the programmer instructs EDAT to consume events matching the provided dependencies. If these dependencies can not be satisfied then task will be paused, the task switched out and the worker free to execute other pending tasks. At some point, when the event dependencies specified by the programmer arrive, the task will be reactivated and a worker will continue to run the task after this line of code. The events themselves are returned as an \emph{EDAT\_Event} structure. Similarly to the submission of tasks, the programmer provides the number of event dependencies and for each of these the source of the event and an event identifier. 

\begin{lstlisting}[frame=lines,caption={Task wait API call},label={lst:taskwait}]
EDAT_Event* edatWait(int number_event_dependencies, <int event_source, char * event_identifier>...)
\end{lstlisting}

This is useful as it enables the programmer to start a task with a minimal set of dependencies met, do some work and then wait until other dependencies arrive to do further work with those. The major advantage of this over splitting the task into a number of sub-tasks is that of context, where task's local variables are maintained and can be used as before once the task re-activates. 

Of course there is some overhead associated with the thread level context switch required when pausing a task and as such the EDAT API also provides a \emph{edatRetrieveAny} call which is a variation of the task wait. If events corresponding to the dependencies are not available, instead of pausing the task for them to arrive, the task continues executing regardless and the call also provides the number of events returned. This is useful in situations where other work could be done by the task even if the events have not yet arrived and allows the programmer to obtain a subset of events and process these as they are made available.

\subsection{Safely sharing local data}
\label{sec:sharingdata}
There are most likely multiple workers, such as one per core in a NUMA region, and so multiple tasks per process are likely executing concurrently. A common challenge with this form of concurrency is avoiding conflict, especially if these tasks are working on the same state. There are a number of ways to protect global state and the programmer can use events to enforce ordering and/or mutual exclusion which is shown in Listing \ref{lst:events_protecting_data}. In this example a persistent task is submitted by Process 0 (line 2) to update some state whenever values are received from Process 1. Shared data is allocated which is provided as the data payload to the \emph{edatFireEvent} call at line 5, with the type \emph{EDAT\_ADDRESS} that specifies the programmer is passing a reference to the data rather than the data itself. When the event \emph{values} arrives from Process 1, the task will run on a worker but, regardless of whether other events with the same \emph{value} identifier also arrive from Process 1 whilst this task is running, another copy of the persistent task will not execute as it also relies on the \emph{data} event. Just before the task completes it fires an event with identifier \emph{data} and the same shared data address to itself (line 10). Effectively this ensures that only one instance of the task is running at any one time and maintains mutual exclusion over the shared state.

\begin{lstlisting}[frame=lines,caption={Events to protect local data},label={lst:events_protecting_data},numbers=left, framexleftmargin=1.5em, xleftmargin=2em]
if (edatGetRank() == 0) {
  edatSubmitPersistentTask(task, 2, EDAT_SELF, "data", 1, "values");
  int * shared_data=(int*) malloc(sizeof(int) * 10);
  edatFireEvent(shared_data, EDAT_ADDRESS, 1, EDAT_SELF, "data");
}

void task(EDAT_Event * events, int num_events) {
  ...
  edatFireEvent(events[0].data, EDAT_ADDRESS, 1, EDAT_SELF, "data");
}
\end{lstlisting}

Whilst managing shared data safety via events fits with the general ethos of EDAT, it does represent locking at a rather coarse grained level. The programmer can use \emph{edatWait} to make this more fine grained, or alternatively EDAT provides \emph{edatLock}, \emph{edatUnlock} and \emph{edatTestLock} calls for explicitly acquiring locks. These all accept a string lock name as an identifier and, in addition to explicit unlocking by the programmer, when a task finishes all locks that it has acquired will be automatically released. When the programmer pauses a task using the \emph{edatWait} of Section \ref{sec:pausetasks}, all acquired locks will also be released automatically and when the task reactivates all these locks will be reacquired before task execution continues.

\section{Graph 500 BFS benchmark}
\label{sec:graph500results}
The open-source Graph500 benchmark \cite{murphy2010introducing} focuses on data intensive supercomputing and whilst the computation in the benchmark is trivial, the communication is irregular, unpredictable and the most significant performance factor. In the latest version of the reference benchmark the steering committee have developed an active messaging layer on top of MPI which abstracts the specifics of the underling calls. This is a prime example of how MPI is insufficient for these types of codes and application developers forced to develop their own abstractions. Not only does this approach add considerably to the work of the end-developers, but inevitably not all will be experienced HPC programmers and as such might missing opportunities for optimisation and performance. Whilst the Graph 500 is an open-source code, the active messaging layer is highly tied to this application and as such is not easily reusable by other codes.  

In this paper we are focusing on the breadth first search (BFS) kernel which, starting from a single source vertex, traverses a randomly generated graph in phases to find and label the neighbours of vertices. The kernel proceeds until all neighbours of all vertices are labelled. The reference implementation proceeds in a level synchronous approach, where neighbours of all vertices in a given level of the graph are communicated before the processes explicitly synchronise, unpack received information and then proceed with received neighbour information onto the next level in the graph. %The active messaging layer developed by the Graph 500 committee is heavily optimised around this specific problem and functions to handle the received data are executed sequentially by processes after all neighbours of a given level have been communicated.

\begin{figure}[htb]
\centering\includegraphics[width=0.2\textwidth]{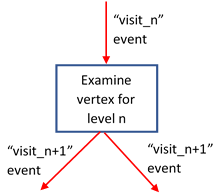}\caption{EDAT BFS task graph}

\label{fig:edatbfstaskgraph}
\end{figure}

Figure \ref{fig:edatbfstaskgraph} illustrates the EDAT task graph for our implementation of the BFS kernel. To compare like with like we implemented a level synchronous approach in EDAT. Even though adding in this extra synchronisation to our EDAT driven BFS kernel is not necessarily the most natural way of using our approach, this avoids revisiting nodes and helps us conduct a fair comparison. A persistent task is scheduled for specific levels in the graph, \emph{n}. This task depends on events with the identifier \emph{visit\_n} (where \emph{n} is the level) and for each vertex will store the required parent information and examine each neighbouring vertex, sending an event to the process holding that neighbour vertex with the identifier \emph{visit\_n+1} where \emph{n+1} is the next level. 

\begin{figure}[htb]
\centering\includegraphics[width=0.45\textwidth]{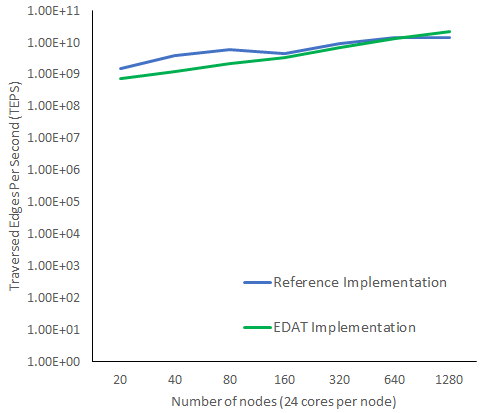}\caption{Graph500 level synchronous BFS benchmark}

\label{fig:graph500results}
\end{figure}

Figure \ref{fig:graph500results} illustrates the performance of of the Graph 500 reference implementation and EDAT version of the BFS kernel (compiled with the GNU 6.3 compiler). This is measured in Traversed Edges Per Second (TEPS) and is for a problem size of scale 29 (536 million verticies) with an edge factor of 16 (8.5 billion edges), run on a Cray XC30. It can be seen that, whilst the performance of the reference version outperforms the EDAT version at lower numbers of nodes, the EDAT version catches up and by 1280 nodes (30720 physical cores) the EDAT version is out performing the reference implementation (2.23e+10 TEPS vs 1.50e+10 TEPS.) 

We found that the best configuration for the EDAT version was to run with one process per physical core, each of these cores containing one worker. The reason behind this is that the visiting task needs to interact with some global state (i.e. whether the vertex has been visited previously and for the scheduling of the next level.) When running with multiple workers then this state needs to be protected via locks (see section \ref{sec:sharingdata}) for every vertex visited. Bearing in mind there are 536 million verticies with this scale problem, continually locking and unlocking adds overhead. In terms of polling for progress we found the optimal setting was to enable the dedicated progress thread and also enable hyperthreading. Effectively each physical core presents itself as two logical cores, one of which was running the progress thread and the other a worker to execute tasks. For a fair comparison with the reference benchmark we also experimented with enabling hyperthreading and the MPI progress thread but this did not significantly improve the performance of that code.

There is some additional overhead associated with the scheduling of tasks and managing of dependencies and this is a major factor behind the lower performance at smaller core counts. However at larger core counts the scalability that EDAT provides becomes more beneficial and, even though we are only running one worker per process and the algorithm is level synchronous, the fact that the progress thread can progress in the background, gather in the events and have these all ready when the task is scheduled is beneficial in contrast to the reference approach. Additionally the reference implementation very much works in steps of communicating neighbours for the next level, gathering verticies for the next level and updating its state. Our implementation, whilst it is still working in a level synchronous manner, combines the update of state and communication of neighbours, effectively running them concurrently, which is of benefit.

%A challenge is that of communication data size and task granularity. For communication data size, EDAT's MPI transport layer automatically coalesces events fired from one process to another in a small time frame. This is entirely hidden from the programmer, but allows the runtime to optimise message sizes between the processes. The other challenge is that of granularity because the computation associated with processing a single \emph{visit} event is tiny and one task per event would take around 1e-7 seconds, the overhead of task scheduling is more significant than this. Instead, in order to amortise the overhead of tasking, the same task can be used to process any number of events using the \emph{edatRetrieveAny} call. This call also accepts the \emph{EDAT\_ANY} constant for the number of events which will return any number of outstanding events matching the specified dependencies. In this manner we can use one task per process per level, continually retrieving the outstanding visit events for that level until level processing has completed.

This benchmark also illustrates that our approach can be applied incrementally to existing codes rather than requiring an entire rewrite that is often needed by other task-based technologies such as Charm++ and Legion. In this work just the BFS kernel itself was rewritten in EDAT, with other part of the Graph500 benchmark such as graph generation, validation and edge counting remaining in MPI. The EDAT library has been written in a way that supports inter-operability with application level MPI which we feel is important because it enables the programmer to concentrate their effort on porting scalability hotspots in their code without having to rewrite en-mass other, less critical, areas. 
% Task granularity
\section{In-situ data analytics case-study}
% Only a few changes, no big modifications needed
\label{sec:casestudy}
The Met Office NERC Cloud model (MONC) \cite{brown2015highly} is an open-source (BSD licensed) high resolution modelling framework that employs large eddy simulation to study the physics of turbulent flows and further develop and test physical parametrisations and assumptions used in numerical weather and climate prediction. MONC is a major model used by UK atmospheric and climate communities, and modern parallelisation has opened up the possibility of running on tens of thousands of cores and soon to be hundreds of thousands. However in addition to generating the raw, \emph{prognostic} data, the model also needs to analyse this into higher level \emph{diagnostic} information. It does this via an in-situ approach \cite{brown2018situ} where cores are split up between computation and data analytics; typically one core in a NUMA region runs the data analytics and \emph{services} the remaining computation cores. 

The MONC developers wrote their own be-spoke analytics code where computational processes send data at different times and in different orders to their corresponding in-situ core. Therefore, whilst the in-situ cores often have to communicate, there is no knowing exactly when each analytics core will be ready to issue this communication or the order of communications; for instance one core might issue communications for field A and then B whilst other cores issue communications for B and then A. The nature of communications typically involves very many small messages, often involving all the analytics cores. It is important to embrace this irregularity, not only because global synchronisation protocols would be expensive, but also to keep each analytics core busy processing data to avoid their computational cores being forced to stall.

\begin{figure}[h]
	\begin{center}    
		\includegraphics[scale=0.30]{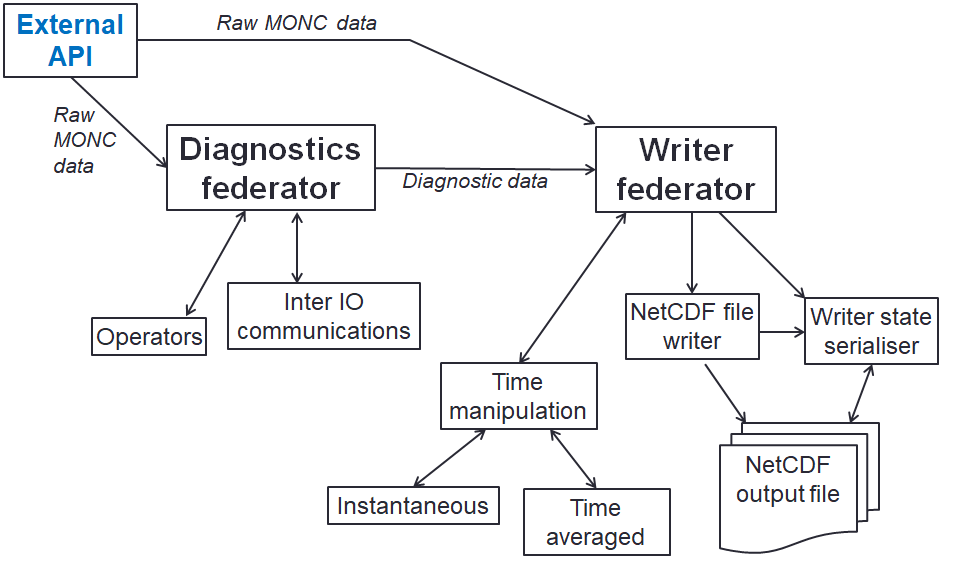}
	\end{center}     
	\caption{MONC in-situ data analytics pipeline}
	\label{fig:moncpipeline}
\end{figure}

The developers of MONC wrote their bespoke communications layer on top of MPI point-to-point for the data analytics to handle the irregular nature of their application. This adds considerable complexity and bloat to the code, is very much tied to MONC and is also a performance bottleneck. Figure \ref{fig:moncpipeline} illustrates the data analytics pipeline of MONC. In the existing MONC code a thread from a pool was activated to handle requests in the \emph{external API} from a computational core such as raw data message arrival. In this case that same thread is then used to process and write the data. Whenever analytics cores need to communicate with each other (e.g. the \emph{inter IO communications}) data is sent tagged with the timestep and field name and the root communication core will start a new thread to process the resulting messages. This mixing in application code of logic and communications mechanism is messy, and the application programmer needs to mix higher and lower level concerns which is not ideal.

The interactions of figure \ref{fig:moncpipeline} are really an instance of the event based co-ordination pattern \cite{mattson2004patterns} and as such, it was our hypothesis that EDAT would be a natural way of expressing this use-case. We replaced the MONC be-spoke communications code with EDAT instead (which has bindings for Fortran). The \emph{external API} submits a single persistent task on start-up for computational core registration and once this happens, via the firing an event matching the dependency of task registration, then two further tasks are submitted, a de-registration task for that core and a persistent task handling data from that core. Also, on start-up, the writer federator and diagnostics federators each submit a persistent task, within the diagnostic federator EDAT tasks are also used for the \emph{inter IO communications}. This has resulted in a decrease of 9\% in the code size of the in-situ data analytics and, most importantly, this code was the tricky, low level logic around communications, rather than the data analytics application-specific code. An additional 2\% of the analytics code was modified, with the remaining 89\% of the code unchanged. This illustrates how our approach can be retrofitted into existing applications without wide scale code changes.

\begin{figure}
	\begin{center}
		\includegraphics[scale=0.50]{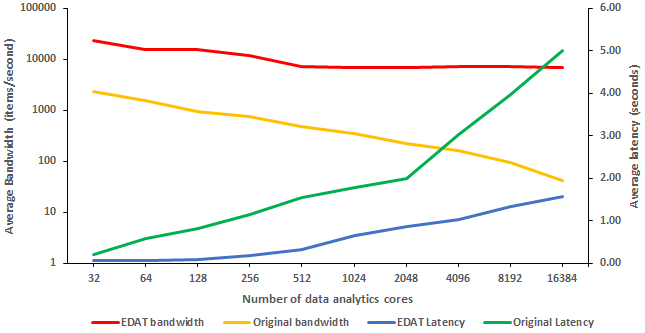}
	\end{center}
	\caption{Bandwidth and latency against number of data analytics cores for EDAT and original MONC in-situ data analytics}
	\label{fig:performanceresults}
\end{figure}

Performance tests on a Cray XC30 (using the GNU 6.3 compiler), have been performed and for the purposes of benchmarking there is a 1 to 1 ratio between computational and data analytics cores. This is so we could run with a large number of analytics cores, all of which need to communicate. A standard, stratus cloud, test case has been used with the computational code modified to repeatedly send raw data, saturating its corresponding data analytics core with as much data as possible. In the EDAT implementation we configured 10 workers per analytics process and in the original MONC code have set the same number of threads. The analytics required for each value involves both arithmetic operations and reductions between all analytics cores. Figure \ref{fig:performanceresults} illustrates the results of these tests, measuring both the average bandwidth (number of items per second that can be processed, ignoring the file write time), and latency (average amount of time it takes for each raw data item to be processed and produce a resulting value, again ignoring the time writing to disk) against the number of data analytics cores. 

From the results of figure \ref{fig:performanceresults} it can be seen that the bandwidth afforded by the EDAT approach is an order of magnitude greater than the original, bespoke, MONC code. The average latency is also significantly smaller for the EDAT implementation in comparison to the original code. One of the major reasons for this is because the MONC developers had not formalised their communications via tasks and there was a complex inter-dependence between threads, and certain situations where progress can not be made without a number of threads running concurrently. Memory cleaning functionality was also required in the original MONC communications code, freeing up the state of previous interactions, which for safety must lock out many other activities running. Instead, by formalising the interaction into tasks the overall architecture is more obvious, poor design choices are avoided, much of the state cleaning is performed by the underlying EDAT library and the diagnostic and writer federators now run concurrently.

% 2 workers per EDAT, MONC threading is 10 threads

Global communication between the analytics cores (a reduction) is required for processing each data item. Hence both the bandwidth decreases and latency increases as the number of analytics cores increases. But also the reduction root is automatically distributed amongst the analytics cores such that activities after communication are more distributed as the number of analytics cores is increased. This explains the levelling off of bandwidth, at around 7000 items per second at larger core counts for the EDAT version. As the original MONC code contains significant overhead it does not benefit from this increased distribution.

\section{Conclusions and further work}

Driven by the rise of areas such as data science and machine learning, there are vast numbers of potential HPC users in communities not traditionally associated with supercomputing, which have the potential to transform our field. However many of these communities have strong open-source roots and an eco-system without open-source software and hardware will likely limit uptake. Additionally, the parallel workloads of these new communities is likely to be far more unpredictable and irregular, along with a large number of programmers who are potentially already familiar with writing asynchronous codes in other domains such as web-development. Current ubiquitous parallel programming technologies are not well suited to these new types of application, and whilst task-based models look like a promising approach, the current generation all exhibit disadvantages for large scale codes.

In this paper we have described a new, open-source, approach to task-based models where the programmer is explicitly aware of the distributed nature of their application but still abstracted from the low level details of how components interact. Splitting problems up into distinct tasks and driving the interaction of these via events, provides a high level view of parallelism. This means programmers can focus on important areas such the design of parallelism, and optimisations including locality. 

The general idea of our approach is simple and few new concepts need to be mastered. It is our opinion that tasks and events maps well to how people should be thinking about parallelism; their parallel tasks, where these are located and how they interact. We have demonstrated good performance and scaling of EDAT with both the Graph500 BFS benchmark on up to 1280 nodes (30720 cores) and the real world MONC model with 16384 data analytics cores. Not only does this illustrates the ability of our approach to scale well to high core counts, but additionally provides an important step change in capability, especially for in-situ data analytics, that these sorts of applications can now take advantage of.

So far we have concentrated on events explicitly fired by user code, but they could represent other behaviours. For example an event could be generated based on some resource becoming available, the activation of a timer, or a hardware event. There could be any number of machine generated events that then allow the programmer to submit tasks that interact with different states of the machine itself in a unified manner. Our approach is also applicable to heterogeneous architectures, for instance EDAT could fire events, potentially directly from the GPU to the NIC without having to go via the CPU and this is also an avenue for further work.

%Additionally there is more work to do around the abstractions for task granuality, we support the \emph{edatRetrieveAny} call with \emph{EDAT\_ANY} to retrieve any number of matching events and we think that task submission could be further extended to also include this in a natural way.

%Our approach is also applicable to heterogeneous architectures and it would be interesting to explore this. For instance many GPU codes require inter-node communication and channelling this through the CPU, often then using MPI to perform the node to node communications, can add considerable complexity and be a bottleneck. Instead, using EDAT to fire events, potentially directly from the GPU to the NIC without having to go via the CPU, could provide a useful tool for programmers. 

The EDAT library implements the concepts described in this paper and has been used for the experiments conducted. EDAT is open-source, BSD licensed and available at https://github.com/EPCCed/edat which also includes documentation, examples and further benchmarks. We are very keen for more contributors and users to become involved and, whilst the library is perfectly usable for real world codes, inevitably there are further library level optimisations that can be made in the future. In addition to being directly usable from C and C++, Fortran and Python bindings are also provided.

\bibliographystyle{IEEEtran}
\bibliography{main.bib}
\end{document}